\documentclass[onecolumn,sort&compress,numbers]{els-mrw-mod}

\usepackage{amsmath,amssymb,amsfonts,amsthm,makeidx,graphicx}
\usepackage{txfonts}
\usepackage{helvet}

\usepackage{bm}
\usepackage{slashed}
\usepackage{axodraw2}
\usepackage{tcolorbox}
\tcbuselibrary{breakable}

\newcommand{\bq}{\begin{eqnarray}}
\newcommand{\eq}{\end{eqnarray}}
\newcommand{\eps}{\varepsilon}
\newcommand{\loopnumber}{l}

\newcommand{\nexternal}{n}

\newcommand{\nedges}{n_{\mathrm{tot}}}
\newcommand{\nvertices}{r_{\mathrm{tot}}}

\newcommand{\ninternal}{n_{\mathrm{int}}}
\newcommand{\ninternalvertices}{r_{\mathrm{int}}}

\newcommand{\ntoy}{N}

\newcommand{\Nmaster}{N_{\mathrm{master}}}
\newcommand{\NL}{N_L}
\newcommand{\NB}{N_B}

\begin{document}

%
%
\chapter{Feynman Diagrams}
\label{chap1}

\author[1]{Stefan Weinzierl}

\address[1]{\orgname{Johannes Gutenberg-Universität Mainz}, \orgdiv{Institut für Physik}, \orgaddress{Staudinger Weg 7, D - 55099 Mainz, Germany}}

\articletag{Chapter Article tagline: update of previous edition, reprint.}

\maketitle

\begin{abstract}[Abstract]
	We give a concise and pedagogical introduction to Feynman diagrams.
        After discussing a toy model which requires only undergraduate mathematics,
        we focus on relativistic quantum field theory.
        We review the derivation of Feynman rules from the Lagrangian of the theory
        and we discuss modern methods to compute tree and loop diagrams.
\end{abstract}

\begin{keywords}
        perturbation theory
        \sep
        precision calculations
        \sep
        Feynman diagrams
        \sep
        Feynman integrals
        \sep
        Feynman rules
\end{keywords}

\begin{figure}[h]
	\centering
	\includegraphics[width=7cm,height=4cm]{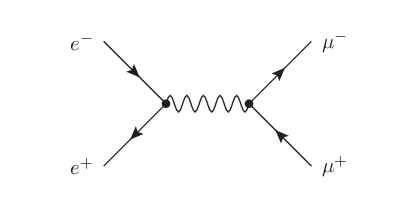}
	\caption{Example of a Feynman diagram contributing to the scattering $e^+ e^- \rightarrow \mu^+ \mu^-$.}
	\label{fig:titlepage}
\end{figure}

\section*{Objectives}
This article answers the following questions:
\begin{itemize}
	\item What is a Feynman diagram and how do they occur in perturbative calculations?
	\item How is a Feynman diagram translated into a mathematical expression?
        \item What are modern methods to compute Feynman diagrams?
\end{itemize}

\section{Introduction}
\label{sect:intro}

Feynman diagrams are pictorial representations of individual terms in the perturbative expansion of a quantity, 
for example a scattering amplitude.
Feynman diagrams have been introduced in 1948 by Richard Feynman 
in the context of quantum electrodynamics \cite{Feynman:1948ur},
but have a much wider application.
They are commonly used in perturbative calculations in relativistic quantum field theory, 
but also in other fields, like solid state physics or gravitational wave physics.
In this article we focus on relativistic quantum field theory, as this theory underlies the Standard Model of particle physics.
Feynman diagrams are a useful tool to organise a perturbative calculation.
In essence, to compute the contribution at a certain order within perturbation theory, 
one first draws all Feynman diagrams contributing at this order and -- in a second step -- translates each diagram into a 
mathematical expression. 
The sum of all these expressions gives then the desired contribution.
The dictionary for the translation from a diagram towards a mathematical expression is given by the Feynman rules.

\section{A toy model}
\label{sect:toy_model}

In order to motivate Feynman diagrams let us consider a simple finite-dimensional toy model which involves only undergraduate mathematics.
Suppose we are interested in $\ntoy$-dimensional integrals of the form
\bq
\label{eq_toy_family}
 I_{\bm \nu}\left(\lambda\right)
 & = &
 \frac{1}{{\bm \nu}!}
 \int d^{\ntoy}\phi
 \; {\bm \phi}^{\bm \nu} \; e^{-\frac{1}{2} {\bm \phi}^T {\bm P} {\bm \phi} - \frac{\lambda}{24} \sum\limits_{i=1}^{\ntoy} \phi_i^4},
\eq
where ${\bm \phi}^{\bm \nu}$ is a short-hand notation for $\phi_1^{\nu_1} \dots \phi_{\ntoy}^{\nu_{\ntoy}}$,
the expression ${\bm \nu}!$ is a short-hand notation for $\nu_1! \dots \nu_{\ntoy}!$,
the $\nu_i$'s are non-negative integers,
${\bm P}$ is a real symmetric positive definite $\ntoy \times \ntoy$ matrix, 
$\lambda$ is a small real non-negative parameter
and the integration is over the components of the real vector ${\bm \phi}=(\phi_1,\dots,\phi_{\ntoy})^T$.
The assumption that ${\bm P}$ is a positive definite matrix implies that ${\bm P}$ is an invertible matrix.
Due to the quartic term in the exponent,
the integral in eq.~(\ref{eq_toy_family}) is difficult to evaluate analytically.
We may resort to perturbation theory and evaluate this integral
as a power series in $\lambda$:
\bq
 I_{\bm \nu}
 & = &
 \sum\limits_{l=0}^\infty I_{\bm \nu}^{(l)},
\eq
where $I_{\bm \nu}^{(l)}$ contains $l$ factors of $\lambda$.
The term $I_{\bm \nu}^{(l)}$ is obtained by expanding 
the exponential $e^{- \frac{\lambda}{24} \sum\limits_{i=1}^{\ntoy} \phi_i^4}$ and by picking the terms of order $\lambda^l$.
Doing so, we have to calculate integrals of the form $I_{\bm \nu'}(0)$.
If we set $|{\bm \nu}|=\nu_1+\dots+\nu_{\ntoy}$, we have for the integrals contributing at order $\lambda^l$
the relation $|{\bm \nu}'| = |{\bm \nu}| + 4l$.
The integrals $I_{\bm \nu'}(0)$ have at most quadratic terms in the exponent and can be computed from Gaussian integrals as follows:
\bq
\label{eq_integral}
 I_{\bm \nu'}\left(0\right)
 \; = \;
 \frac{1}{{\bm \nu}'!}
 \int d^{\ntoy}\phi
 \; {\bm \phi}^{{\bm \nu}'} \; e^{-\frac{1}{2} {\bm \phi}^T {\bm P} {\bm \phi}}
 \; = \; 
 \frac{1}{{\bm \nu}'!}
 \left. \left[
 \prod\limits_{i=1}^{\ntoy} \left(\frac{\partial}{\partial J_i}\right)^{\nu_i'}
 \int\limits
 d^{\ntoy}\phi  \; e^{-\frac{1}{2} {\bm \phi}^T {\bm P} {\bm \phi} + {\bm J}^T {\bm \phi}}
 \right] \right|_{{\bm J}={\bm 0}}
 \; = \;
 \frac{\left( 2 \pi \right)^{\frac{n}{2}}}{{\bm \nu}'! \sqrt{\det {\bm P} }}
 \left. \left[
 \prod\limits_{i=1}^{\ntoy} \left(\frac{\partial}{\partial J_i}\right)^{\nu_i'}
 e^{\frac{1}{2} {\bm J}^T {\bm P}^{-1} {\bm J}}
 \right] \right|_{{\bm J}={\bm 0}}.
\eq
In principle, we could stop here as we have reduced the problem of computing perturbatively 
the integral in eq.~(\ref{eq_toy_family}) to differentiation.
In order to see how Feynman diagrams emerge, we press on and ask what terms are obtained by the differentiation
on the right-hand side of eq.~(\ref{eq_integral}).
It is easy to see that terms with an odd number of differentiations vanish by setting ${\bm J}={\bm 0}$ in the end.
For an even number of differentiations
the analogue of Wick's theorem \cite{Wick:1950ee} adapted to our toy example reads
\bq
\label{eq_Wick}
 \left. \left[
 \frac{\partial}{\partial J_{i_1}} \dots \frac{\partial}{\partial J_{i_{2m}}}
 e^{\frac{1}{2} {\bm J}^T {\bm P}^{-1} {\bm J}}
 \right] \right|_{{\bm J}={\bm 0}}
 & = &
 \sum\limits_{\mathrm{pairings}}
 \left({\bm P}^{-1}\right)_{i_{\sigma_1} i_{\sigma_2}}
 \dots
 \left({\bm P}^{-1}\right)_{i_{\sigma_{2m-1}} i_{\sigma_{2m}}},
\eq
where the sum is over all possibilities of grouping the elements of $\{i_1,\dots,i_{2m}\}$ into sets of two.
In eq.~(\ref{eq_Wick}) it is not required that the elements of $\{i_1,\dots,i_{2m}\}$ are pairwise distinct.
We represent $({\bm P}^{-1})_{i j}$ by a line segment, connecting the endpoints $i$ and $j$:
\bq
\label{toy_propagator}
\begin{axopicture}(50,15)(0,25)
\Line(10,30)(40,30)
\Vertex(10,30){2}
\Vertex(40,30){2}
\Text(7,25)[rt]{$i$}
\Text(43,25)[lt]{$j$}
\end{axopicture}
 & = &
 \left({\bm P}^{-1}\right)_{i j}.
\eq
Let us now consider the perturbative expansion of $I_{\bm \nu}(\lambda)/I_{\bm 0}(0)$ for
${\bm \nu} = (1,1,0,0,\dots,0)^T$.
At order $\lambda^0$ we only have $({\bm P}^{-1})_{1 2}$, which can represented by a line segment as in eq.~(\ref{toy_propagator}).
At order $\lambda^1$ we obtain
\bq
\label{example_order_1}
 \frac{I_{\bm \nu}^{(1)}}{I_{\bm 0}\left(0\right)}
 & = &
 - \lambda
 \sum\limits_{i=1}^{\ntoy}
 \left[ 
   \frac{1}{2} \left({\bm P}^{-1}\right)_{1 i} \left({\bm P}^{-1}\right)_{i i} \left({\bm P}^{-1}\right)_{i 2}
 + \frac{1}{8} \left({\bm P}^{-1}\right)_{1 2} \left({\bm P}^{-1}\right)_{i i}^2
 \right].
\eq
Graphically, we represent this as
\bq
 \frac{I_{\bm \nu}^{(1)}}{I_{\bm 0}\left(0\right)}
 & = &
\left(
\begin{axopicture}(90,30)(0,25)
\Line(10,20)(70,20)
\Vertex(10,20){2}
\Vertex(40,20){2}
\Vertex(70,20){2}
\CArc(40,30)(10,0,360)
\Text(7,15)[rt]{$1$}
\Text(40,15)[t]{$i$}
\Text(73,15)[lt]{$2$}
\end{axopicture}
\right)
 \;\;\;
+
 \;\;\;
 \left(
\begin{axopicture}(90,30)(0,25)
\Line(10,30)(40,30)
\Vertex(10,30){2}
\Vertex(40,30){2}
\Vertex(70,30){2}
\CArc(70,40)(10,0,360)
\CArc(70,20)(10,0,360)
\Text(7,25)[rt]{$1$}
\Text(43,25)[lt]{$2$}
\Text(80,30)[l]{$i$}
\end{axopicture}
\right),
\eq
with the following conventions: We call the vertices $1$ and $2$ external vertices, the vertex $i$
an internal vertex.
For each internal vertex $i$ we include a factor $(-\lambda)$ and sum over $i$ from $1$ to $\ntoy$.
Each external vertex contributes a trivial factor $1$.
As above, an 
edge between vertices $i$ and $j$ contributes a factor $({\bm P}^{-1})_{i j}$.
Finally, we divide each graph by the order of the permutation group
of the internal vertices and edges leaving the diagram unchanged when the external vertices are fixed.
The order of the permutation group is also called the symmetry factor of the diagram.
The symmetry factor of the first diagram equals $2$, as we may interchange the two ends of the edge $({\bm P}^{-1})_{i i}$.
The symmetry factor of the second diagram equals $8$: We may interchange the two edges $({\bm P}^{-1})_{i i}$, in addition
we may interchange for each edge $({\bm P}^{-1})_{i i}$ the two endpoints.
Note that a graph needs not to be connected. Note further that the sum over $i$ includes the cases $i=1$ and $i=2$,
hence we don't draw the latter graphs separately.
The above conventions for translating a graph into a mathematical formula are called the Feynman rules for our toy example.

Let us return to our original integral of eq.~(\ref{eq_toy_family}).
Of particular interest is the case $\nu_j \in \{0,1\}$.
In this case we obtain the order $\lambda^l$ contribution to $I_{\bm \nu}(\lambda)/I_{\bm 0}(0)$ as follows:
Draw all possible diagrams with one external $1$-valent vertex for each $\nu_j=1$
and $l$ internal $4$-valent vertices.
Translate each diagram to a mathematical expression with the help of the Feynman rules.
The sum of these expressions gives the contribution $\lambda^l I_{\bm \nu}^{(l)}/I_{\bm 0}(0)$.

Note that the Feynman rule for an edge is determined by the terms in the exponent, which are quadratic in ${\bm \phi}$, while
the Feynman rule for an internal vertex is determined by the term in the exponent, which is quartic in ${\bm \phi}$.

In order to appreciate the advantages of Feynman diagrams you are invited to compute by hand the right-hand side of eq.(\ref{example_order_1}) in two ways:
Once by differentiation with the help of eq.~(\ref{eq_integral})
and once with the help of the Feynman rules given above.

\section{Feynman rules}
\label{sect:rules}

Let us now turn to relativistic quantum field theory.
The above toy example captures already many features of Feynman diagrams and Feynman rules.
The main generalisation is to go from a finite number of degrees of freedom to an infinite number by replacing
$\phi_i$ by $\phi_i(x)$, e.g. $\ntoy$ degrees of freedom for every space-time point $x$.
As the number of space-time points is uncountable, so are the degrees of freedom.
We denote the number of space-time dimensions by $D$ and we assume that the theory is specified by
a Lagrangian ${\mathcal L}({\bm \phi})$.
The Lagrangian may contain derivatives acting on ${\bm \phi}$.
It is customary to set the speed of light and Planck's constant to one: $c=\hbar=1$.
Integrating the Lagrangian over space-time yields the action:
\bq
\label{def_action}
 S\left[{\bm \phi} \right] & = & \int d^Dx \; {\mathcal L}\left({\bm \phi} \right).
\eq
The analogue of eq.~(\ref{eq_toy_family}) is
\bq
\label{eq_path_integral}
 \frac{1}{{\bm \nu}!}
 \int {\mathcal D} {\bm \phi}
 \; {\bm \phi}^{\bm \nu} \; e^{i S[{\bm \phi}]},
\eq
where only a finite number of the $\nu_i(x)$'s are non-zero, say $\nexternal$ of them.
In the following we are interested in the case where $\nu_i(x) \in \{0,1\}$.
In this case, $\frac{1}{{\bm \nu}!} {\bm \phi}^{\bm \nu}$ reduces to
${\bm \phi}(x_1) \dots {\bm \phi}(x_{\nexternal})$.
The integral over all field configurations in eq.~(\ref{eq_path_integral})
is an infinite-dimensional integral and is called a path integral.
The toy example of section~\ref{sect:toy_model} can be considered a specific example in $D=0$ space-time dimensions.

The theory is interacting, if the the Lagrangian contains terms cubic or higher in the fields.
In this case it is difficult to evaluate the path integral exactly.
If all interaction terms come with a small coupling, we may use perturbation theory.
This is explained in detail in almost all textbooks of quantum field theory, see for example \cite{Peskin,Schwartz,Srednicki:2007qs}.
The central object is the generating functional defined as
\bq
 Z[{\bm J]} & = & {\mathcal N} \int {\mathcal D} {\bm \phi} \; e^{i \left( S[{\bm \phi}] + \int d^Dx \; {\bm J}(x) {\bm \phi}(x) \right)}.
\eq
The prefactor ${\mathcal N}$ is chosen such that $Z[0]=1$.
In our toy example we encountered the analogue of the generating functional in eq.~(\ref{eq_integral}).
The $\nexternal$-point Green function is given by
\bq
 \langle 0| T( {\bm \phi}(x_{1}) ... {\bm \phi}(x_{\nexternal})) |0 \rangle
 & = &
 {\mathcal N} 
 \int {\mathcal D} {\bm \phi} \; {\bm \phi}(x_{1}) ... {\bm \phi}(x_{\nexternal}) \; e^{i S({\bm \phi})}.
\eq
With the help of functional derivatives this can be expressed as
\bq
 \langle 0| T( {\bm \phi}(x_{1}) ... {\bm \phi}(x_{\nexternal})) |0 \rangle & = &
   \left. \left(-i \right)^{\nexternal}
   \frac{\delta^{\nexternal} Z[{\bm J}]}{\delta {\bm J}(x_1) ... \delta {\bm J}(x_{\nexternal})} \right|_{{\bm J}=0}.
\eq
In analogy with section~\ref{sect:toy_model} one may work out the Feynman rules for the 
$\nexternal$-point Green functions and compute the latter perturbatively from Feynman diagrams.
In practice, we are not so much concerned with the $\nexternal$-point Green functions.
The main interest are the scattering amplitudes.
These are obtained through three modifications:
(i) one only considers connected diagrams,
(ii) one goes from position space to momentum space and
(iii) one considers amputated Green functions.
Technically, this is done as follows:
The connected Green functions in position space are obtained from a functional $W[{\bm J}]$, which is
related to the functional $Z[{\bm J}]$ by $Z[{\bm J}] = e^{i W[{\bm J}]}$.
The connected Green functions are then given by
\bq
 G_{\nexternal}\left(x_1,...,x_{\nexternal}\right) 
 & = & 
 \left( -i \right)^{{\nexternal}-1} 
 \left. \frac{\delta^{\nexternal} W[{\bm J}]}{\delta {\bm J}(x_1) ... \delta {\bm J}(x_{\nexternal}) } \right|_{{\bm J}=0}.
\eq
The transition from position space to momentum space is done by a Fourier transformation.
One defines the Green functions in momentum space by
\bq
\label{eq_fourier_trafo}
 G_{\nexternal}\left(x_1,...,x_{\nexternal}\right) 
 & = &  
 \int \frac{d^Dp_1}{(2\pi)^D} ... \frac{d^Dp_{\nexternal}}{(2\pi)^D}
 e^{-i \sum\limits_{j=1}^{\nexternal} p_j x_j} \left(2 \pi \right)^D \delta^D\left(p_1+...+p_{\nexternal}\right) 
 \tilde{G}_{\nexternal}\left(p_1,...,p_{\nexternal}\right).
\eq
Note that the Fourier transform $\tilde{G}_{\nexternal}$ is defined by explicitly factoring out the 
$\delta$-distribution $\delta^D(p_1+...+p_{\nexternal})$ and a factor $(2 \pi )^D$.
We denote the two-point function in momentum space by $\tilde{G}_2(p)$.
In this case we have to specify only one momentum,
since the momentum flowing into the Green function on one side has to be equal to the momentum flowing
out of the Green function on the other side due to the presence of the $\delta$-distribution in eq.~(\ref{eq_fourier_trafo}) .
We may now define scattering amplitudes: In momentum space a 
scattering amplitude with $\nexternal$ external particles is
given by the connected $\nexternal$-point Green function multiplied by the
inverse two-point function for each external
particle:
\bq
 i {\mathcal A}_{\nexternal}\left(p_1,...,p_{\nexternal}\right)
 & = &
 \tilde{G}_2\left(p_1\right)^{-1}
 ...
 \tilde{G}_2\left(p_{\nexternal}\right)^{-1}
 \tilde{G}_{\nexternal}\left(p_1,...,p_{\nexternal}\right).
\eq
The multiplication with the inverse two-point function for each external
particle amputates the external propagators.
As a consequence we distinguish in a graph external and internal edges.

The valency of a vertex in a graph is the number of edges attached to it.
Vertices of valency $0$, $1$ and $2$ are special.
A vertex of valency $0$ is necessarily disconnected from the rest of graph and therefore not relevant for connected graphs.
A vertex of valency $1$ has exactly one edge attached to it. 
This edge is called an external edge. 
All other edges are called internal edges.
In the context of scattering amplitudes it is common practice not to draw a vertex of valency 1, but just the external edge.
A vertex of valency $2$ is also called a dot.
In the context of scalar loop integrals dots are often used to indicate that a propagator occurs to a power higher than one.
In the context of Feynman diagrams the use of the word ``vertex'' usually implies a vertex of valency $3$ or greater.
This derives from the fact that in a particle picture a vertex of valency $3$ or greater corresponds
to a genuine interaction among particles.
The loop number of a connected graph with 
$\ninternal$ internal edges and 
$\ninternalvertices$ internal vertices
is defined to be
\bq
 \loopnumber & = & \ninternal-\ninternalvertices+1.
\eq
The graph in fig.~\ref{fig:titlepage} has one internal edge and two internal vertices, hence the loop number
is $\loopnumber = 1-2+1=0$.
The definition of the loop number is adapted to the case where we don't draw external vertices.
For a scattering amplitude with $\nexternal$ particles
the total number of edges is $\nedges=\ninternal+\nexternal$
and the total number of vertices is $\nvertices=\ninternalvertices+\nexternal$
(including the external vertices which we usually don't draw), hence we also have
$\loopnumber = \nedges-\nvertices+1$, which is the standard definition in the mathematical literature.
The loop number has the following interpretation:
If we fix all momenta of the external lines and if we impose momentum conservation at each internal vertex, 
then the loop number is equal to the number of
independent momentum vectors not constrained by momentum conservation.
A connected graph of loop number $0$ is called a tree graph.
Connected graphs with loop number $\loopnumber > 0$ are called loop graphs.

From the Lagrangian of the theory we obtain the Feynman rules for this theory.
The most important Feynman rules are the ones for the propagators and the vertices.
These are obtained as follows:
We first order the terms in the Lagrangian according to the number of fields they involve.
From the terms bilinear in the fields one obtains the propagators,
while the terms with three or more fields give rise to vertices.
Note that a standard Lagrangian does not contain terms with just one or zero fields.
A term without any field is just an irrelevant constant, while terms with just one field indicate that quantum fluctuations
are not around an extremal value of the action.
Furthermore we always assume within perturbation theory that all fields fall off rapidly enough at infinity.
Therefore we can use partial integration and ignore boundary terms.
A generic bilinear term for a real field $\phi_i$ can therefore be cast in the form
\bq
 {\mathcal L}_{\mathrm{bilinear}}
 & = &
 \frac{1}{2}
 \sum\limits_{i,j}
 \phi_i(x) P_{ij}(x) \phi_j(x),
\eq
where $P$ is a real symmetric operator that may contain derivatives and must have an inverse. 
Define the inverse of $P$ by
\bq
 \sum\limits_j P_{ij}(x) P_{jk}^{-1}(x-y) & = & \delta_{ik} \delta^D(x-y),
\eq
and its Fourier transform by
\bq
 P_{ij}^{-1}(x) & = & \int \frac{d^D q}{(2 \pi)^D} e^{-i q \cdot x} \tilde{P}_{ij}^{-1}(q).
\eq
Then the Feynman rule for the propagator is given by
\bq
\label{Feynman_rule_propagator}
 \Delta_F(q)_{ij} 
 & = & 
 i \tilde{P}_{ij}^{-1}(q).
\eq
Let us consider a few examples: $\phi^4$-theory is specified by the Lagrangian
\bq
{\mathcal L}^{\mathrm{scalar}}
 & = & 
 \frac{1}{2} \phi(x) \left[ -\Box - m^2 \right] \phi(x) + \frac{1}{4!} \lambda \phi(x)^4,
\eq
where we denoted the d'Alembert operator by $\Box=\partial_\mu \partial^\mu$.
The first term is bilinear in the fields and we have $P(x)=-\Box - m^2$.
The propagator for a scalar particle is represented by a line and given by
\bq
 \begin{axopicture}(75,20)(0,5)
 \Line(20,10)(70,10)
\end{axopicture} 
 \;\; = \;\;
 \frac{i}{q^2-m^2}.
\eq
As a second example we consider quantum electrodynamics with 
the covariantly gauge-fixed Lagrangian 
\bq
 {\mathcal L}^{\mathrm{QED}}
 & = &
 \frac{1}{2} A_{\mu}(x) \left[ \partial_\rho \partial^\rho g^{\mu\nu}
                  - \left( 1 - \frac{1}{\xi} \right) \partial^\mu \partial^\nu \right] A_{\nu}(x)
 + \overline{\psi}(x) \left(  i \gamma^{\, \mu} \partial_{\mu} - m \right) \psi(x)
 + e \; \overline{\psi}(x) \; \gamma^{\mu} A_{\mu}(x) \; \psi(x),
\eq
where $A_\mu(x)$ denotes the photon field, 
$\psi(x)$ denotes the electron field, $\gamma^\mu$ the Dirac matrices 
and $e$ the absolute value of the electron charge.
We also used Einstein's summation convention.
We first consider the photon propagator. From the terms 
bilinear in the photon fields $A_\mu(x)$ we extract
\bq
 P^{\mu\nu}(x) & = & \partial_\rho \partial^\rho g^{\mu\nu}
                  - \left( 1 - \frac{1}{\xi} \right) \partial^\mu \partial^\nu.
\eq
For the propagator we are interested in the inverse of this operator
\bq
 P^{\mu\sigma}(x) \left( P^{-1} \right)_{\sigma\nu}(x-y) & = & g^\mu_{\;\;\nu} \delta^D(x-y).
\eq
Working in momentum space we are more specifically interested in the Fourier transform of the inverse of this operator:
\bq
 \left( P^{-1} \right)_{\mu\nu}(x) & = & 
  \int \frac{d^D q}{(2 \pi)^D} e^{-i q \cdot x} \left( \tilde{P}^{-1} \right)_{\mu\nu}(q).
\eq
The Feynman rule for the propagator is then given by $(\tilde{P}^{-1})_{\mu\nu}(q)$ times the imaginary unit.
For the photon propagator one finds the Feynman rule
\bq
 \begin{axopicture}(75,20)(0,5)
 \Photon(20,10)(70,10){-5}{4.5}
 \Text(15,12)[r]{\footnotesize $\mu$}
 \Text(75,12)[l]{\footnotesize $\nu$}
\end{axopicture} 
 & = & 
  \frac{i}{q^2} \left( - g_{\mu\nu} + \left( 1 -\xi \right) \frac{q_\mu q_\nu}{q^2} \right).
\eq
In order to distinguish different particles, it is customary to draw edges in a Feynman diagram
for different particles in different ways. A wavy line is the standard convention to represent a photon.
In a similar way one may derive the electron propagator.
One finds
\bq
 \begin{axopicture}(75,20)(0,5)
 \ArrowLine(20,10)(70,10)
\end{axopicture} 
 \;\; = \;\;
 \frac{i\left(\slashed{q}+m\right)}{q^2-m^2}.
\eq
The momentum $q$ is flowing in the direction of the arrow.

Let us now turn to the vertices and let us consider a generic interaction term with $n\ge 3$ fields.
We may write this term as
\bq
 {\mathcal L}_{\mathrm{int}} & = & 
 \sum\limits_{i_1 \dots i_n}
 O_{i_1 \dots i_n}\left(\partial_1, \dots ,\partial_n \right)
 \phi_{i_1}(x) \dots \phi_{i_n}(x),
\eq
with the notation that $\partial_j$ acts only on the $j$-th field $\phi_{i_j}(x)$.
For each field we have the Fourier transform
\bq
 \phi_i(x) = 
 \int \frac{d^Dq}{(2\pi)^D} \; e^{-i q x} \; \tilde{\phi}_i(q),
 & &
 \tilde{\phi}_i(q) =
 \int d^Dx \; e^{i q x} \; \phi_i(x),
\eq
where $q$ denotes an in-coming momentum.
We thus have
\bq
 {\mathcal L}_{\mathrm{int}} & = & 
 \int
 \frac{d^Dq_1}{(2\pi)^D} \dots \frac{d^Dq_n}{(2\pi)^D}
 e^{-i \left(q_1+\dots+q_n\right) x}
 O_{i_1 \dots i_n}\left(-i q_1,\dots,-i q_n \right)
 \tilde{\phi}_{i_1}\left(q_1\right) \dots \tilde{\phi}_{i_n}\left(q_n\right).
\eq
Changing to outgoing momenta we replace $q_j$ by $-q_j$.
The vertex is then given by
\bq
\label{feynman_rule_vertex}
 V & = & 
 i \sum\limits_{\mathrm{permutations}} 
   (-1)^{P_F} O_{i_1 \dots i_n}\left(i q_1,\dots,i q_n \right),
\eq
where the momenta are taken to flow outward.
The summation is over all permutations of indices and momenta of identical particles. 
In the case of identical fermions there is in addition a minus sign for every odd permutation 
of the fermions, indicated by $(-1)^{P_F}$.

Let us also look at some examples.
We start again with scalar $\phi^4$-theory.
There is only one interaction term,
containing four field $\phi(x)$:
\bq
 {\mathcal L}^{\mathrm{scalar}}_{\mathrm{int}}
 & = & 
 \frac{\lambda}{4!}
 \phi(x) \phi(x) \phi(x) \phi(x).
\eq
Thus $O = \lambda/4!$ and the Feynman rule for the vertex is given by
\bq
\begin{axopicture}(100,20)(0,50)
\Vertex(50,50){2}
\Line(50,50)(66,66)
\Line(50,50)(66,34)
\Line(50,50)(34,34)
\Line(50,50)(34,66)
\end{axopicture}
 & = & i \lambda. 
 \\
 \nonumber 
\eq
The factor $1/4!$ is cancelled by summing over the $4!$ permutations of the four identical particles.

As a second example we consider the 
interaction term in quantum electrodynamics between electrons and photons, given by
\bq
 {\mathcal L}^{\mathrm{QED}}_{\mathrm{int}}
 & = & 
 e \; \overline{\psi}(x) \; \gamma^{\mu} A_{\mu}(x) \; \psi(x),
\eq
where $\psi(x)$ denotes the electron field and $e$ the absolute value of the electron charge.
We have
\bq
 O^\mu\left(i q_1, i q_2 ,i q_3 \right)
 & = &
 e \gamma^{\mu} 
\eq
and the Feynman rule
\bq
\begin{axopicture}(100,15)(0,55)
\Vertex(50,50){2}
\Photon(50,50)(80,50){3}{4}
\ArrowLine(50,50)(29,71)
\ArrowLine(29,29)(50,50)
\Text(82,50)[lc]{$\mu$}
\end{axopicture}
 & = & i e \gamma^{\, \mu}.
 \\
 \nonumber \\ 
 \nonumber
\eq
If all couplings describing the strengths of interactions among the particles are small,
we may calculate scattering amplitudes within perturbation theory.
Let us assume for simplicity that there is only one coupling, which we denote by $g$
(in the example of quantum electrodynamics we have $g=e$, in the example 
of $\phi^4$ theory we set $g=\sqrt{\lambda}$).
We expand the scattering amplitude in powers of $g$:
\bq
\label{basic_perturbative_expansion}
 {\mathcal A}_{\nexternal} 
 & = & 
 {\mathcal A}_{\nexternal}^{(0)} + {\mathcal A}_{\nexternal}^{(1)} + {\mathcal A}_{\nexternal}^{(2)} + {\mathcal A}_{\nexternal}^{(3)} + \dots,
\eq
where ${\mathcal A}_{\nexternal}^{(\loopnumber)}$ contains $({\nexternal}-2+2\loopnumber)$ factors of $g$.
Eq.~(\ref{basic_perturbative_expansion}) gives the perturbative expansion of the
scattering amplitude.
In this expansion, ${\mathcal A}_{\nexternal}^{(\loopnumber)}$ is an amplitude with ${\nexternal}$ external particles and $\loopnumber$ loops.
In order to compute ${\mathcal A}_{\nexternal}^{(\loopnumber)}$ from Feynman diagrams
we first draw all Feynman diagrams for the given number of external particles $\nexternal$ and the given number of loops $\loopnumber$. 
We then translate each graph into a mathematical formula with the help of the Feynman rules given below.
The quantity $i {\mathcal A}_{\nexternal}^{(\loopnumber)}$ is then given as the sum of all these terms.
\begin{tcolorbox}[breakable]
Feynman rules for scattering amplitudes:
\begin{enumerate}

\item For each internal edge include a propagator according to eq.~(\ref{Feynman_rule_propagator}).

\item For each internal vertex we have momentum conservation at this vertex. Include a factor according to eq.~(\ref{feynman_rule_vertex}).
\item For each external edge include a factor describing the spin polarisation of the corresponding particle.
Thus, there is a polarisation vector $\eps^\mu(q)$ for each external spin-$1$ boson and a 
spinor $\bar{u}(q)$, $u(q)$, $\bar{v}(q)$ or $v(q)$ for each external spin-$\frac{1}{2}$ fermion.
For spin-$0$ bosons there is no non-trivial spin polarisation to be described, hence 
an external edge corresponding to a spin-$0$ boson translates to the trivial factor $1$.

\item There is an integration
\bq
 \int \frac{d^Dq}{(2\pi)^D}
 \nonumber
\eq
for each internal momentum not constrained by momentum conservation.

\item Each diagram is multiplied by a factor $1/S$, where $S$ is the order of the permutation group
of the internal lines and vertices leaving the diagram unchanged when the external lines are fixed.

\item A factor $(-1)$ for each closed fermion loop.

\end{enumerate}
\end{tcolorbox}
For each edge we choose an orientation, indicating the momentum flow through this edge. 
Momentum conservation at a vertex is the statement, that the sum of all momenta flowing into this vertex equals the sum of all momenta
flowing out of this vertex.
Note that the loop integration corresponds in the toy model of section~\ref{sect:toy_model} to the summation
over the internal index $i$.
The minus sign for each closed fermion loop follows from the anti-commutation rules for fermionic operators.

\section{Tree diagrams}
\label{sect:trees}

Tree-level amplitudes ${\mathcal A}_{\nexternal}^{(0)}$ are computed from Feynman diagrams with no loops.
They give the leading contribution to the full amplitude.
The computation of tree-level amplitudes involves only basic mathematical operations like
addition, multiplication or the contraction of indices.
Thus tree-level amplitudes can therefore be computed in principle for any number of external particles.
In practice, the number of contributing Feynman diagrams is a limiting factor for tree-level amplitudes
with a large number of external particles. 

As an example we consider the tree-level amplitude for the process $e^-e^+ \rightarrow \mu^- \mu^+$ in quantum electrodynamics.
For simplicity we take the electrons and the muons to be massless.
There is only one Feynman diagram contributing to this tree-level amplitude, this is the diagram shown in fig.~\ref{fig:titlepage}.
It is convenient to take all momenta outgoing, thus we compute
\bq
 0 \rightarrow \mu^-(p_1) \; \mu^+(p_2) \; e^-(p_3) \; e^+(p_4).
\eq
An outgoing particle with momentum $p$ equals an ingoing anti-particle with momentum $(-p)$.
Momentum conservation reads then $p_1+p_2+p_3+p_4=0$.
Applying the Feynman rules to the diagram of fig.~\ref{fig:titlepage} yields with $p_{12}=p_1+p_2$
\bq
 {\mathcal A}_{4}^{(0)}
 & = &
 \bar{u}(p_1) i e \gamma^\mu v(p_2) 
 \frac{i}{p_{12}^2} \left( - g_{\mu\nu} + \left( 1 -\xi \right) \frac{p_{12\, \mu} p_{12\, \nu}}{q^2} \right)
 \bar{u}(p_3) i e \gamma^\nu v(p_4) 
 \; = \;
 \frac{ie^2}{p_{12}^2} 
 \bar{u}(p_1) \gamma^\mu v(p_2) 
 \bar{u}(p_3) \gamma_\mu v(p_4).
\eq
At the last equal sign we used the Dirac equation for massless particles $\bar{u}(p) \slashed{p} = \slashed{p} v(p) = 0$.

As already mentioned above, the number of contributing Feynman diagrams is a 
limiting factor for (tree-level) amplitudes with a large number of external particles. 
There are several techniques, which improve the efficiency of a calculation.
These techniques are explained in detail in several
excellent review articles and lecture notes \cite{Mangano:1990by,Dixon:1996wi,Elvang:2013cua,Weinzierl:2016bus,Badger:2023eqz}.
Here, we summarise the main ideas.

For the cross-section we need the scattering amplitude squared, summed over all internal degrees of freedom like spin or colour.
Assume that ${\mathcal A}_{\nexternal}^{(0)}$ is given as the sum of $N_{\mathrm{terms}}$ terms.
$N_{\mathrm{terms}}$ can be a rather large number.
Squaring the amplitude and summing over the spins will result in ${\mathcal O}(N_{\mathrm{terms}}^2)$ terms.
It is more efficient to use explicit representations for the polarisation states, 
calculate the amplitude for each polarisation configuration (this is an ${\mathcal O}(N_{\mathrm{terms}})$ operation),
square the amplitude (for a given polarisation configuration this is an ${\mathcal O}(1)$ operation, basically computing the norm of a complex number)
and then summing over all polarisation configurations.
If each of the $\nexternal$ external particles has two polarisation states, the computational cost is
\bq
 2^{\nexternal} N_{\mathrm{terms}},
\eq
which for large $N_{\mathrm{terms}}$ is much smaller than $N_{\mathrm{terms}}^2$.
This is the idea of the spinor helicity method \cite{DeCausmaecker:1981jtq,Xu:1986xb}.
In practice one even avoids the prefactor $2^{\nexternal}$ by Monte Carlo sampling over the helicity configurations
or a Monte Carlo integration over helicity angles \cite{Draggiotis:1998gr}.

Let us compute one helicity amplitude for the process discussed above: We consider the helicity configuration 
$(\lambda_1,\lambda_2,\lambda_3\lambda_4) = (+,-,+,-)$.
As we assume all particles to be massless, it is convenient to work with two-component Weyl spinors $\langle p \pm |$ and $| p \pm \rangle$.
An outgoing fermion with positive helicity is represented by the Weyl spinor $\langle p+ |$, 
an outgoing anti-fermion with negative helicity by the Weyl spinor $| p+ \rangle$.
Introducing the spinor products $\langle i j \rangle = \langle p_i- | p_j+ \rangle$, $[ i j ] = \langle p_i+ | p_j- \rangle$ 
and using the Fierz identity one obtains
\bq
 {\mathcal A}_{4}^{(0)}\left(p_1^+, p_2^- p_3^+ p_4^-\right)
 & = &
 \frac{2 ie^2}{p_{12}^2} 
 [ 1 3 ] \langle 4 2 \rangle. 
\eq
In this particular example one can show that all other tree-level helicity amplitudes are either zero or related by symmetry to the one above.

A second efficiency improvement is colour decomposition. 
Scattering amplitudes in quantum chromodynamics 
(or more generally any scattering amplitude involving gauge bosons related to an unbroken gauge symmetry)
may be decomposed into group-theoretical factors (carrying the colour structures)
multiplied by kinematic functions called partial amplitudes \cite{Cvitanovic:1980bu,Berends:1987cv,Mangano:1987xk}. These partial amplitudes
do not contain any colour information and are gauge-invariant objects. 
As an example let us look at gluon scattering at tree-level, i.e. the process $0 \rightarrow g_1 \dots g_{\nexternal}$.
The colour decomposition reads in this case
\bq
 {\mathcal A}_{\nexternal}^{(0)}\left(p_1,p_2,...,p_{\nexternal}\right) 
 & = & g^{\nexternal-2} \sum\limits_{\sigma \in S_{\nexternal}/Z_{\nexternal}} 
 2 \; \mathrm{Tr} \left( T^{a_{\sigma(1)}} ... T^{a_{\sigma(\nexternal)}} \right)
 \;\;
 A_{\nexternal}^{(0)}\left( p_{\sigma(1)}, ..., p_{\sigma(\nexternal)} \right),
\eq
where the sum is over all non-cyclic permutations of $\{1,2,...,\nexternal\}$.
The quantities $A_{\nexternal}^{(0)}( p_{\sigma(1)}, ..., p_{\sigma(\nexternal)} )$ accompanying
the colour factor $2 \; \mathrm{Tr}( T^{a_{\sigma(1)}} ... T^{a_{\sigma(\nexternal)}})$
are called partial amplitudes or primitive amplitudes.
For tree-level Yang-Mills amplitudes the notions of partial amplitudes and primitive amplitudes
coincide, although this is no longer true if one considers amplitudes with quarks and/or amplitudes with loops.
The primitive amplitudes for tree-level Yang-Mills amplitudes are calculated 
from planar diagrams with a fixed cyclic ordering of the external legs
and cyclic-ordered Feynman rules.
There are two simplifications occurring: First of all, the 
cyclic-ordered Feynman rules are simpler than the standard Feynman rules, in particular the four-gluon vertex expands to fewer terms.
Secondly, due to the restriction of cyclic-ordering, there are fewer diagrams contributing to a primitive amplitude.
Table~\ref{table_number_of_diagrams} shows a comparison of the number of diagrams contributing to 
the full amplitude ${\mathcal A}_{\nexternal}^{(0)}$
\begin{table}
\begin{center}
\bq
\begin{array}{l|rrrrrrr}
 n                    & 4 & 5  & 6   & 7    & 8     & 9      & 10 \\
 \hline
\mbox{unordered}      & 4 & 25 & 220 & 2485 & 34300 & 559405 & 10525900 \\
\mbox{cyclic ordered} & 3 & 10 & 38  & 154  & 654   & 2871   & 12925 \\
\end{array}
\eq
\caption{\label{table_number_of_diagrams} The number of diagrams contributing to the full amplitude ${\mathcal A}_{\nexternal}^{(0)}$
and to the cyclic-ordered primitive amplitude $A_{\nexternal}^{(0)}$ for the process $0 \rightarrow g_1 \dots g_{\nexternal}$ at tree-level.}
\end{center}
\end{table}
and to the cyclic-ordered primitive amplitude $A_{\nexternal}^{(0)}$.

The spinor-helicity method and colour decomposition deal with the internal degrees of freedom of spin and colour, respectively.
It remains to calculate efficiently a primitive helicity amplitude.
Naively summing up all Feynman diagrams one is still hampered 
by the large number of contributing Feynman diagrams
even for a moderately large number of external particles.

For numerical calculations, off-shell recurrence relations provide an efficient way to compute primitive helicity amplitudes.
Off-shell recurrence relations \cite{Berends:1987me}
build primitive helicity amplitudes from smaller building blocks, called cyclic-ordered off-shell currents.
Off-shell currents are objects with $j$ on-shell legs and one additional leg off-shell, with $j$ ranging from $1$ 
to $(\nexternal-1)$.
Momentum conservation is satisfied. It should be noted that
off-shell currents within gauge theories are not gauge-invariant objects.
Recurrence relations relate off-shell currents with $j$ legs 
to off-shell currents with fewer legs.
The computation of a primitive helicity amplitude based
on off-shell recurrence relation scales polynomially with the number of external particles.
The most efficient algorithms achieve a scaling of $\nexternal^3$.
This polynomial behaviour is much better than the factorial growth of an algorithm based on Feynman diagrams.
The re-use of the results for the already computed lower-point currents is essential in achieving this
polynomial behaviour.

For analytic calculations, on-shell recurrence relations \cite{Britto:2004ap,Britto:2005fq}
provide an efficient way to compute primitive helicity amplitudes.
On-shell recursion relations compute the primitive tree amplitude $A_{\nexternal}^{(0)}$ with ${\nexternal}$
external particles recursively from primitive tree amplitudes with fewer legs.
Obviously, the primitive tree amplitudes with fewer legs entering this calculation satisfy momentum conservation
and the on-shell conditions.
Furthermore, one deals in the case of gauge theories at all times with gauge-invariant quantities.
However, at each step in the recursion, the external momenta of the primitive tree amplitudes with fewer legs
are redefined and in general will take complex values.
The redefinition of the momenta slows down a numerical computation of primitive helicity amplitudes based on on-shell
recurrence relations, and off-shell recurrence relations are the better option for a numerical computation.
On-shell recursion relations are the method of choice for compact analytical formulae.

\section{Loop diagrams}
\label{sect:loops}

Tree-diagrams give usually the leading order in the perturbative expansion.
For precision calculations one would like to include more orders in the perturbative expansion.
In this case loop diagrams enter the game
and the Feynman rules involve an integration over each internal momentum not constrained
by momentum conservation.
Textbooks on Feynman integrals are \cite{Smirnov:2012gma,Weinzierl:2022eaz}.

Let us start with a very simple loop diagram, the one-loop tadpole with $(\nu-1)$ dots in $D$ space-time dimensions (with $\nu \in {\mathbb N}$).
\bq
\label{def_tadpole}
 T_\nu\left(D,m^2\right)
 & = &
 \int \frac{d^Dk}{\left(2\pi\right)^D} 
 \frac{1}{\left(k^2-m^2\right)^\nu}.
\eq
We face two basic problems:
The first problem is the correct integration contour.
Let us assume that space-time has the Lorentzian metric $g_{\mu\nu}=\mathrm{diag}(1,-1,-1,-1,\dots)$
and let us write $k=(k^0,\vec{k})$ for the $D$-dimensional loop-momentum vector with one energy component $k^0$ 
and a spatial $(D-1)$-dimensional momentum vector $\vec{k}$.
As a function of $k^0$, the integrand has poles at $k^0=\pm\sqrt{\vec{k}^2+m^2}$
and we have to specify how to treat these singularities of the integrand.
The correct prescription follows from causality in quantum field theory and instructs us to avoid the pole
at $k^0=\sqrt{\vec{k}^2+m^2}$ by escaping into the complex upper half-plane, 
and to avoid the pole
at $k^0=-\sqrt{\vec{k}^2+m^2}$ by escaping into the complex lower half-plane.
An equivalent formulation is to replace the denominator $k^2-m^2$ by $k^2-m^2+i\delta$, where $\delta$ is infinitesimal.
This is Feynman's $i\delta$-prescription and applies to all propagators in a Feynman integral.
Very often the small imaginary parts are not written explicitly, but they are always understood implicitly.

To see the second problem we consider the massless tadpole with $\nu=2$ in four space-time dimensions,
i.e. eq.~(\ref{def_tadpole}) for $D=4$, $m=0$ and $\nu=2$.
After Wick rotation to Euclidean space, the introduction of spherical coordinates in four dimensions 
and integration over the three angular variables one arrives at
\bq
 T_2\left(4,0\right)
 & = &
 \int \frac{d^4k}{\left(2\pi\right)^4} 
 \frac{1}{\left(k^2\right)^2}
 \; = \;
 \frac{i}{16 \pi^2}
 \int\limits_0^\infty \frac{dx}{x}. 
\eq
This integral diverges at $x\rightarrow \infty$, which is called an 
ultraviolet divergence and at $x\rightarrow 0$, which is called an infrared divergence.
Any quantity, which is given by a divergent integral, is of course an ill-defined quantity.
Therefore the first step is to make these integrals well-defined by introducing a regulator.
Within perturbative quantum field theory
the method of dimensional regularisation \cite{tHooft:1972tcz,Bollini:1972ui,Cicuta:1972jf}
has almost become a standard, as the calculations in this regularisation
scheme turn out to be the simplest.
Within dimensional regularisation one replaces the four-dimensional integral over the loop momentum by a
$D$-dimensional integral, where $D$ is now an additional parameter, which can be a non-integer or
even a complex number.
We consider the result of the integration as a function of $D$ and we are interested in the behaviour of this 
function as $D$ approaches $4$.
Anticipating dimensional regularisation we already started in eq.~(\ref{def_action}) with $D$ space-time dimensions.
Dimensional regularisation works for the following reason:
By using spherical coordinates and Feynman or Schwinger parameters, 
we may always ensure that the integrand of any Feynman integral is independent of the angular variables. 
The integration over the $(D-1)$ angles yields then
\bq
\label{angular_integration}
 \int\limits_{0}^{\pi} d\theta_{1} \sin^{D-2} \theta_{1}
 ... \int\limits_{0}^{\pi} d\theta_{D-2} \sin \theta_{D-2} 
 \int\limits_{0}^{2 \pi} d\theta_{D-1} 
 & = & \frac{2 \pi^{\frac{D}{2}}}{\Gamma\left( \frac{D}{2} \right)},
\eq
where $\Gamma(z)$ denotes Euler's gamma function.
On the right-hand side of eq.~(\ref{angular_integration}) we no longer need to restrict to $D \in {\mathbb N}$,
but may view the expression as a function of $D \in {\mathbb C}$.
 
Within dimensional regularisation we consider the integral in $D=4-2\eps$ space-time dimensions
and we expand the result as a Laurent series in the dimensional regularisation parameter $\eps$.
The divergences in four space-time dimensions show up as the pole terms in the Laurent expansion.
Poles originating from ultraviolet divergences are removed by renormalisation, 
poles originating from infrared divergences cancel 
due to the Kinoshita-Lee-Nauenberg theorem \cite{Kinoshita:1962ur,Lee:1964is}
when summed over all degenerate physical states.

The integrand of a Feynman integral in the momentum representation is always a rational function in the
the loop momenta.
The denominator of this rational function is given by a product of factors $(q_j^2-m_j^2)$, where
$q_j$ denotes the momentum flowing through edge $e_j$ and $m_j$ denotes the mass of the particle propagating along edge $e_j$.
If the numerator of this rational function is independent of all loop momenta, we call the Feynman integral a scalar integral,
otherwise a tensor integral.
From the Feynman rules for most quantum field theories
(like Yang-Mills theory, QED, QCD or more generally any quantum field theory, which is not a scalar theory)
we get Feynman integrals, which are tensor integrals.
However, it is always possible to express any tensor integral
as a linear combination of scalar integrals \cite{Tarasov:1996br,Tarasov:1997kx}.
Hence, it is sufficient to focus on scalar integrals.

The scalar Feynman integral 
corresponding to a Feynman graph $G$ with $\nexternal$ external edges, $\ninternal$ internal edges and $\loopnumber$ loops
is given in $D$ space-time dimensions up to prefactors by
\bq
\label{def_scalar_Feynman_integral}
 I_{\nu_1,\dots,\nu_{\ninternal}}\left(D,x\right)
 & = &
 \int \prod\limits_{r=1}^{\loopnumber} \frac{d^Dk_r}{\left(2\pi\right)^D} 
 \prod\limits_{j=1}^{\ninternal} \frac{1}{\left(q_j^2-m_j^2\right)^{\nu_j}},
\eq
where each internal edge $e_j$ of the graph is associated with a triple $(q_j,m_j,\nu_j)$,
specifying the momentum $q_j$ flowing through this edge,
the mass $m_j$ and the power $\nu_j$ to which the propagator occurs.
In the following we will assume $\nu_j \in {\mathbb Z}$.
The external momenta are labelled by $p_1,\dots,p_{\nexternal}$.
Each internal momentum $q_j$ can be written as a linear combination of the $l$ loop momenta $k_r$ 
and the linear independent external momenta $p_s$.
The Feynman integral in eq.~(\ref{def_scalar_Feynman_integral}) depends
on the dimension of space-time $D \in {\mathbb C}$, the $\ninternal$-tuple $(\nu_1,\dots,\nu_{\ninternal})$
and on kinematic variables.
The Feynman integral in eq.~(\ref{def_scalar_Feynman_integral}) is a scalar integral, thus the dependence
on the linear independent external momenta is only through the Lorentz invariants
$p_i \cdot p_j$.
Thus the kinematic variables are the Lorentz invariants $p_i \cdot p_j$ and the internal masses squared $m_j^2$.
We denote the kinematic variables by $x=(x_1, \dots, x_{\NB})$.

At the time of writing of this review article, one of the most powerful methods
to compute Feynman integrals is the method 
of differential equations \cite{Kotikov:1990kg,Kotikov:1991pm,Remiddi:1997ny,Gehrmann:1999as}.
Without loss of generality we may assume that the graph $G$ is such that any Lorentz invariant $k_i \cdot k_j$ or $k_i \cdot p_j$ 
can always be written as a linear combination of the $\ninternal$ propagators $(q_s^2-m_s^2)$.
If a given graph $G$ does not have this property, one can always find a larger graph $\tilde{G}$ with this property and
such that the original graph $G$ is recovered through pinching of some internal edges of $\tilde{G}$.

We now consider a family of Feynman integrals, which share the same $D$ and $x$, 
but differ by the $\ninternal$-tuples $(\nu_1,\dots,\nu_{\ninternal}) \in {\mathbb Z}^{\ninternal}$.
This is a countable set of Feynman integrals.
However, there are linear relations among the members of a family \cite{Tkachov:1981wb,Chetyrkin:1981qh}.
These relations are obtained from integration-by-parts 
identities and based on the fact that within dimensional regularisation the integral
of a total derivative vanishes
\bq
\label{basic_ibp_relation}
 \int 
 \frac{d^Dk}{\left(2 \pi\right)^D}
 \;\;
 \frac{\partial}{\partial k^\mu} \; \left[ q^\mu \cdot f\left(k\right) \right]
 & = & 0,
\eq
i.e. there are no boundary terms. 
The vector $q$ can be any linear combination of the external momenta and the loop momentum $k$.
Applying this to the integral in eq.~(\ref{def_scalar_Feynman_integral}) and letting the derivative act on the integrand
will yield a linear relation among different members of a family.
We may introduce an ordering criteria for the $\ninternal$-tuple $(\nu_1,\dots,\nu_{\ninternal})$
and use the linear relations to eliminate more complicated Feynman integrals 
(with respect to the ordering criteria)
in favour of simpler Feynman integrals \cite{Laporta:2000dsw}.
The ones which cannot be eliminated are called master integrals.
It can be shown that the number of master integrals is finite \cite{Smirnov:2010hn}.
Let us denote the master integrals by
$I_{{\bm{\nu}}_1}, I_{{\bm{\nu}}_2}, \dots, I_{{\bm{\nu}}_{\Nmaster}}$
with ${\bm{\nu}}_i = ( \nu_{i 1}, \dots, \nu_{i \ninternal} )$.
Let us then pick one master integral $I_{{\bm{\nu}}_i}$ and one kinematic variable $x_k$.
By carrying out the differentiation under the integral sign and by using integration-by-parts identities
we may express the derivative $\frac{\partial I_{{\bm{\nu}}_i}}{\partial x_k}$ as a linear combination
of the master integrals.
Doing so for any master integral and any kinematic variable we obtain a system of first order
differential equations for $I=(I_{{\bm{\nu}}_1}, \dots, I_{{\bm{\nu}}_{\Nmaster}})^T$.
\bq
\label{differential_equation}
 d I
 & = &
 A\left(\eps,x\right) I,
\eq
where $A(\eps,x)$ is a $\Nmaster \times \Nmaster$-matrix, whose entries are differential one-forms, rational
in $\eps$ and $x$ and $d=\sum\limits_{j=1}^{\NB} dx_j \frac{\partial}{\partial x_j}$.
It is worth mentioning that there are no conceptional issues in obtaining the differential equation, as it involves only linear algebra.
However, there can be practical problems, if the size of the linear system gets too large.

In addition to the system of differential equation we need boundary values for the master integrals.
The boundary conditions can be trivialised at the expense of possibly introducing additional kinematic variables
in the form of auxiliary masses \cite{Liu:2017jxz,Liu:2022mfb}.

From this point onwards, we may either follow a numerical approach and solve the 
system of differential equations with appropriate boundary values numerically \cite{Liu:2022chg},
or continue with an analytic calculation.
In the latter case one tries to find a new basis of master integrals $I'$, 
related to the old basis of master integrals $I$ by $I'=U I$, with
an $\Nmaster \times \Nmaster$-matrix $U(\eps,x)$ such that
\bq
\label{eps_factorised}
 d I'
 & = &
 \eps A'\left(x\right) I',
\eq
i.e. the only dependence of $\eps A'(x)$ on the dimensional regularisation parameter $\eps$ is through the explicit prefactor \cite{Henn:2013pwa}.
A system of differential equations of this form is said to be $\eps$-factorised.
One writes
\bq
 A'\left(x\right) & = & 
 \sum\limits_{j=1}^{\NL} C_j \; \omega_j\left(x\right),
\eq
where the $C_j$'s are constant (i.e. $x$-independent) $\Nmaster \times \Nmaster$ matrices and the $\omega_j$'s are differential one forms.
The $\omega_j$'s are called letters and the set $\{\omega_1,\dots,\omega_{\NL}\}$ is called the alphabet for this family of Feynman integrals.
An $\eps$-factorised system of differential equations can be solved in a straightforward way in terms of Chen's iterated integrals \cite{Chen}.
In the simplest cases one finds that all letters are of the form $\omega_j=d\ln(r_j(x))$, where $r_j(x)$ is a rational function of $x$.
In this case the family of Feynman integrals can be expressed in terms of multiple polylogarithms \cite{Goncharov_no_note,Borwein:1999js}, i.e. iterated integrals 
with letters
\bq
 \omega_j & = & \frac{dx_{k_j}}{x_{k_j}-c_j}.
\eq
Multiple polylogarithms are explicitly defined by
\bq
 G(\underbrace{0,\dots,0}_{r-\mathrm{times}};y)
 & = & 
 \frac{1}{r!} \ln^r\left(y\right),
 \nonumber \\
 G\left(c_1,c_2\dots,c_r;y\right)
 & = &
 \int\limits_0^y
 \frac{dy_1}{y_1-c_1}
 G\left(c_2\dots,c_r;y_1\right),
 \;\;\;\;\;\;
 \left(c_1, \dots, c_r \right) \; \neq \; \left(0, \dots, 0\right).
\eq
The first line of the definition includes the trivial case $G(;y)=1$.
Multiple polylogarithms can be evaluated for all (complex) values of the arguments \cite{Vollinga:2004sn}.
They correspond to iterated integrals on the moduli space of a curve of genus zero with marked points \cite{Brown:2006}.

Note that not all Feynman integrals can be expressed in terms of multiple polylogarithms. 
The simplest Feynman integral which goes beyond multiple polylogarithms
is the equal-mass two-loop sunrise integral, which relates to a curve of genus one.
Further generalisations include curves of higher genus or geometries of higher dimension.
The most prominent examples for the latter case are Calabi-Yau geometries.
At the time of writing of this article it is a topic of current research to 
characterise the functions to which more complicated Feynman integrals
evaluate \cite{Bourjaily:2022bwx}.

Let us summarise the workflow for loop contributions to scattering amplitudes:
One first expresses all tensor integrals in terms of scalar integrals.
The scalar integrals are then expressed as linear combinations of master integrals with the help of integration-by-parts relations.
It remains to compute the master integrals.
One of the most powerful methods to compute master integrals is based on 
differential equations and boundary values.

\section{Conclusions}
\label{sect:conclusions}
Feynman diagrams are omnipresent in perturbative calculations.
They represent individual terms in the perturbative expansion of a quantity.
In this article we first discussed Feynman diagrams in the context of a simple toy model.
We then focused on relativistic quantum field theory.
We reviewed the derivation of Feynman rules from the Lagrangian of the theory
and we discussed modern methods to compute tree and loop diagrams.

Of course, much more can be said about Feynman diagrams and the interested reader is invited
to consult the references \cite{Mangano:1990by,Dixon:1996wi,Elvang:2013cua,Weinzierl:2016bus,Badger:2023eqz,Smirnov:2012gma,Weinzierl:2022eaz}.
Due to page limitations a selection had to be made in this review.
The modern methods for tree and loop diagrams presented in this review are in the opinion of the author 
currently the most efficient ones to evaluate (generic) Feynman diagrams in relativistic quantum field theory.

\seealso{An introduction to gauge theories and quantum field theory, Perturbation theory}

\bibliographystyle{Numbered-Style}
\bibliography{encyclopedia}

\end{document}